\documentclass{llncs}
\usepackage[utf8]{inputenc}
\usepackage[T1]{fontenc}
\usepackage{graphicx} 

\usepackage{subfig}

\DeclareFontFamily{OT1}{pzc}{}
\DeclareFontShape{OT1}{pzc}{m}{it}{<-> s * [1.10] pzcmi7t}{}
\DeclareMathAlphabet{\mathpzc}{OT1}{pzc}{m}{it}

\usepackage{tikz}
\usetikzlibrary{fit}
\tikzset{%
  highlight/.style={rectangle,rounded corners,fill=red!15,draw,
    fill opacity=0.5,thick,inner sep=0pt}
}

\usepackage{xparse}
\usepackage{tikz}
\usetikzlibrary{matrix,backgrounds}
\pgfdeclarelayer{myback}
\pgfsetlayers{myback,background,main}

\tikzset{mycolor/.style = {line width=1bp,color=#1}}%
\tikzset{myfillcolor/.style = {draw,fill=#1}}%

\NewDocumentCommand{\highlight}{O{blue!40} m m}{%
\draw[mycolor=#1] (#2.north west)rectangle (#3.south east);
}

\NewDocumentCommand{\fhighlight}{O{blue!40} m m}{%
\draw[myfillcolor=#1] (#2.north west)rectangle (#3.south east);
}

\newtheorem{principle}{Design Principle}

\begin{document}

\title{Application of Bitcoin Data-Structures \& Design Principles to Supply Chain Management}
\author{S. Matthew English, Ehsan Nezhadian}



\institute{
Rheinische Friedrich-Wilhelms-Universit{\"a}t Bonn\\
\email{\{english, nezhadian\}@cs.uni-bonn.de}
}

\maketitle              

\begin{abstract}
Heretofore the concept of ``blockchain'' has not been precisely defined. 
Accordingly the potential useful applications of this technology have been largely inflated. 
This work sidesteps the question of what constitutes a blockchain as such and focuses on the architectural components of the Bitcoin cryptocurrency, insofar as possible, in isolation. 
We consider common problems inherent in the design of effective supply chain management systems.
With each identified problem we propose a solution that utilizes one or more component aspects of Bitcoin. 
This culminates in five design principles for increased efficiency in supply chain management systems through the application of incentive mechanisms and data structures native to the Bitcoin cryptocurrency protocol. 
\end{abstract}
\section{Introduction}
Bitcoin is an emergent phenomenon realized through the subtle interaction of multiple data structures and incentive mechanisms. 
In isolation the various components of the Bitcoin technology ecosystem are well known and in some cases have existed for years.

The novelty of Bitcoin was to combine these elements in a previously unimagined way. 
The success of Bitcoin as a cryptocurrency has generated interest in the underlying design principles of the cryptocurrency.
This in turn has prompted some to critically reassess traditional methods used to process information. 
The purpose being to determine the extent to which architectural aspects of Bitcoin might be leveraged to reduce or eliminate current inefficiencies. 

\begin{figure}[!tbp]
\centering
\subfloat[]{\includegraphics[width=0.5\textwidth]{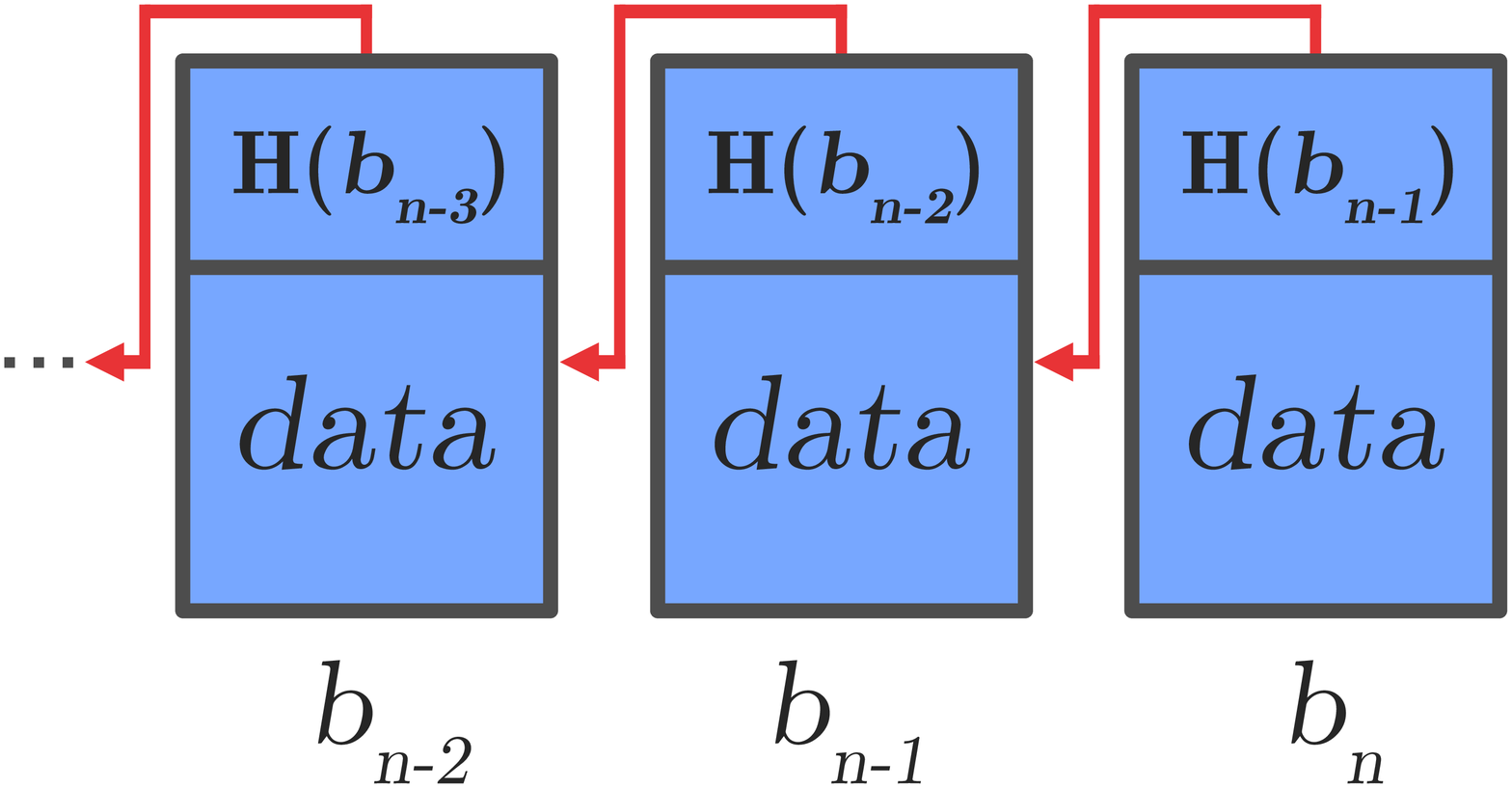}} 
\subfloat[]{\includegraphics[width=0.5\textwidth]{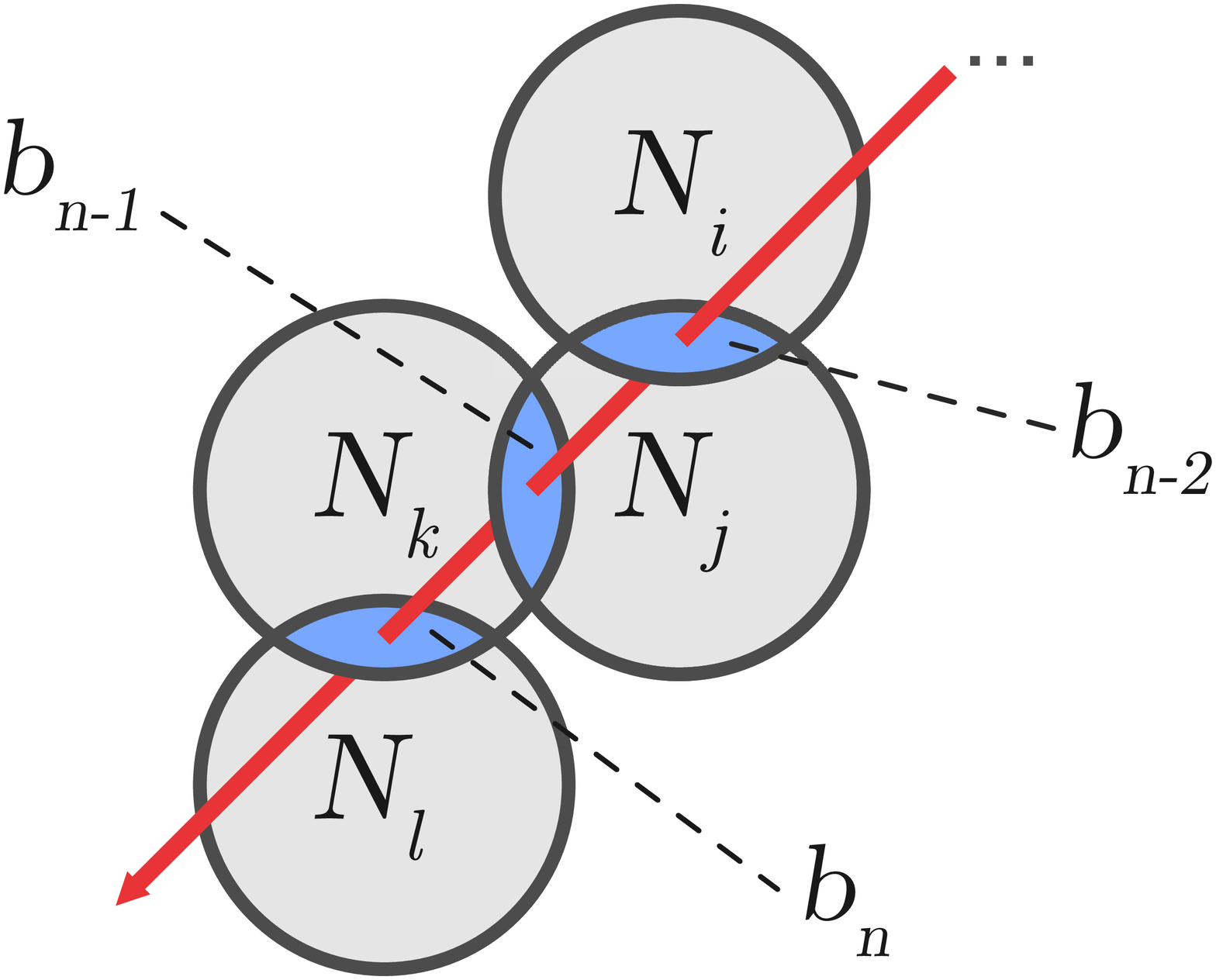}}
\caption{(a) Diagrammatic depiction of a basic blockchain data structure. (b) Representation of a simple transaction flow. The chain proceeds from node $N_i$ (top right) to node $N_l$ (bottom left). Constituent blocks are indicated as $b_i$.
 -- Blue indicates mutual agreement, sealed by proof-of-work (e.g. signature) and committed to the shared record. Nodes not directly connected expose no de-anonymizing information.} 
\label{fig:EcUND} 
\end{figure}

One of the architectural components of Bitcoin is a modified linked list known as a \textit{blockchain}, demonstrated in Figure \ref{fig:EcUND}.
At a fundamental level a blockchain can be thought of as nothing more than a linear collection of data elements, i.e. nodes ($N_1, ..., N_m$). 
Each node, $N_i$, is pointed to by the subsequent node, $N_{i+1}$, through a reference to its hash.
Therefore $N_{i+1}$ maintains a hash of $N_i$. 
One of the characteristics of this data representation format is that the integrity of the complete list can be easily verified with relatively low storage requirements, in fact by maintaining only the single hash at the head of the list.
This construct, introduced in 1990 by Haber \& Stornetta \cite{haber1990time}, is integral to the Bitcoin specification.
A blockchain can be used for verifiably representing and persistently storing information related to supply chains in manufacturing.
However as noted earlier the innovation of Bitcoin is not due to any one element but rather to the interplay of many technical and non-technical components.
In this work we examine a subset of these components and detail a methodology for utilizing them towards the creation of an efficient supply chain management system. 

\section{Related Work} 

The process by which data pertaining to particular parties is safely shared in environments of low trust, such as exists between organizations cooperating in a multi-node supply chain network can be problematic.
Prior efforts to apply the broadly defined concept of ``blockchain'' towards the creation of efficiency gains in supply chain management systems have emphasized the use of distributed computing environments.
The scripting language of Bitcoin is limited in the sense that it is not Turing complete.
Various altcoin implementations have endeavored to provide that functionality.
The purpose of which being the fashioning of a decentralized virtual machine.
Such a system would constitute a distributed app-engine capable of executing programs in a peer-to-peer network.
In order to operate effectively the system would seek to prevent the execution of programs that could be detrimental to the network.
The fundamental issue that operations of this nature are grappling with is the halting problem. 
Operators of such platforms must answer the question of how to ensure that a program eventually terminates and does not waste network resources.
One workaround employed to bypass this impediment has been to rent computation cycles for a fee \cite{wood2014ethereum}. 

To date there have been initial endeavors to apply such systems to the improvement of supply chain management information flows. 
The contributions of these efforts to the creation of a viable solution have thus far been found wanting.
Provenance\footnote{\texttt{https://www.provenance.org/whitepaper}}, one such attempt, states explicitly in the first paragraph of their technical whitepaper ``\textit{the decentralized application (Dapp) proposed in this paper is still in development}'', and goes on to furnish no precise technical specifications.
Skuchain\footnote{\texttt{https://www.skuchain.com}} likewise provides no technical information describing the implementation of their proposed system.
Distinct from these commercial operations, the use of distributed app-engines towards the improvement of supply chain data transmission has been examined in \cite{kim2016towards}, which introduces an ontology-driven model for tracing provenance of goods. 

The relative immaturity of decentralized application systems, such as those employed by the endeavors referenced above, presents a compelling reason to exclude them from a studies such as those that deal with practical applications realizable with current technology. 
In this work we dispense with the notion of employing decentralized application systems. 
We examine exclusively the characteristics inherent to the Bitcoin protocol at the current state of the art.
This paper is intended to provide concrete recommendations for the practical amelioration of supply chain management systems that are realizable with modern engineering and technological capabilities.

One of the most vulnerable attack vectors in Bitcoin protocol inspired supply chain management systems is the mapping from the ``real world'' to the digital world.
The application of RFID to facilitate this inter-linkage as described by \cite{tian2016agri} takes the approach followed here in its identification of tangible inefficiencies and proposal for an actionable corrective measure. 







\section{Data-Structures \& Design Principles}

The exclusive attribution of effects from various properties native to the Bitcoin cryptocurrency protocol is not straightforward as the validity of the system emerges from a complex balance of forces. 
There exists a multi-faceted interchange of data representations and game theoretic incentive mechanisms that give rise to the transaction network which supports Bitcoin. 
Accordingly this section does not attempt to exhaustively enumerate the attributes of that protocol. 
What follows is a high level characterization of a subset of the properties that are inherent to Bitcoin.
Those properties are examined in light of their potential utility to information retention and transmission contextualized in the environment of a multi-party (inter-organizaional) supply chain.  
The selected aspects of the protocol are evaluated in order to provide a mapping between their relevance to Bitcoin and their utility to a supply chain management system. 

\subsection{Pseudo-Anonymity}
Anonymity (and pseudo-anonymity) remains one of the fundamental tenants of the ``cypherpunk'' movement that gave rise to the concept of cryptocurency.
The related property of fungability with respect to individual bitcoins is important to the long-term viability of this technology as a medium of exchange. 
Association with nefarious activity including coins that have been involved in deep-web drug deals, such as those that regularly take place on AlphaBay Market, would potentially subject those coins to censure by authorities, were they to be positively identified. 
Such a result as that just described could diminish public confidence in the currency and precipitate the ultimate dissolution of the transaction network.
What makes cryptocurrencies attractive to counterparties in these illicit transactions is the relative size of the anonymity set associated with their use. 
Network participants are represented only by their address, which if used in accordance with best security practices can be difficult to associate with a real-world identity. 
This serves to emphasize the point that pseudo-anonymity of users is a characteristic regarded by some as indispensable to Bitcoin. 

The practices employed by drug dealers of the $21^{st}$ century, detailed above, stand in stark contrast to the techniques employed by drug dealers of previous eras, such as the 1980's.
The 2001 film Blow presents a portrayal of this reality whilst simultaneously serving to illustrate an important aspect of supply chain management. 
The scene wherein the American cocaine importer, George Jung, introduces his Colombian connection to the head of his California distribution network initiates a process by which George is subsequently extricated from this commercial pipeline. 
This brief anecdote will underscore the importance of anonymity amongst nodes in a supply chain.
Inability to identify nodes to whom one is not directly connected is a critical feature of a supply chain management system, lest that information be used to ``cut out'' intermediary nodes. 
However, representing nodes in a supply chain in such a way that we are able to trace the goods they move through the network is important for preserving provenance of entities exchanged.
Strict prohibition on the de-anonymization of nodes with whom one shares no common edge is simultaneously balanced with the need to trace the provenance of interchanged information. 
There is an analogy in this trade off to the operational characteristics of the Bitcoin protocol. 
Accordingly the pseudo-anonymity properties of the Bitcoin network can be usefully employed in the creation of a shared data structure amongst supply chain nodes with the specified constraints.

\begin{principle}
Nodes must be represented by means of a persistent pseudonymous identifier through which it is possible to associate them with the information exchanged ($t_x$) at an approximate time ($h$). This representation should be resistant to attempts at the association of such nodes with ``real world'' identifying information.
\end{principle}

\subsubsection{Data Replication}

The use of data replication to prevent against the deterioration of information resources in systems requiring a high degree of fault tolerance has been shown to be effective.
This technique has been employed successfully in the distributed processing of large data sets across clusters of computers, such as Hadoop which utilizes redundant copies of information to prevent against unintentional data corruption. 
Replication in computing involves sharing information (distribution) so as to ensure consistency between redundant records.
This process often employs a rudimentary consensus mechanism to establish the canonical source in the event of malfunction, explored later in more detail.  
The Bitcoin protocol utilizes the concept of data replication in the historical transaction data maintained by all full nodes on the network. 
There is a clear analogy here to the individual cluster members in the Hadoop architecture, many of which maintain a distinct copy of (often a subset over) the database.
In the context of a supply chain this feature would enable each node to independently verify their own copy of the flow of goods or merchandise throughout the network.

\begin{principle}
Multiple copies of the shared database, including associated blocks and component transactions, should be maintained by actors operating in the supply chain network exceeding a predetermined threshold.
\end{principle}
 
\subsubsection{Distributed Consensus}

The consensus model in it's most elemental instantiation would take the form of simple majority as in the case of Hadoop detailed above. 
Mutual agreement between counter-parties provides another form of consensus whereby the definitive characterization of an exchange can be recorded after it's having taken place\footnote{The case whereby parties cannot come to an agreement, treated in Section ~\ref{sec:greetings}}. 
Representation of individual nodes in the supply chain by a unique digital signature provides a mechanism for the manufacturing of trust.
If two transacting nodes can consent to affix their personal signatures to a transaction we can consider that this process has been successfully concluded. 
Nodes have permission only to ascent to transactions in which they are directly involved. 
This procedure is roughly sketched in Figure 1. 
Such a process tends to result in a large common history.
The signature take the place of ``proof-of-work'' in the form of reputation staked on the veracity of their participation in an atomic transaction.
If an individual node were to itself become corrupt we rely on the integrity of the remaining network participants that maintain a replicated copy of that transaction. 
The quorum necessary to officiate a particular interpretation can be based on simple majority of a fixed percentage (e.g. 3$\mathpzc{f}$+1)) to establish an implicit consensus on the canonical database representation.

This conception of consensus is dissimilar to the Bitcoin network where trust is a function of the hashing power that a node is able to control. 
The threat model herein considered differs markedly from that necessitated by the maintenance of a distributed cryptocurrency network since it is assumed that nodes exchanging goods in a supply chain already foster a modicum of trust, i.e. a working business relationship.
This assumption permits of more flexibility in the optimal behaviour policy we can expect from nodes. 

\begin{principle}
Consensus is established by ascent of mutually transacting nodes. These atomic instances are committed to a common shared history replicated throughout a subset of the network which serves as the future canonical representation based on a predetermined consent parameter (network percentage).
\end{principle}

\begin{figure}
  \centering
    \includegraphics[width=0.8\textwidth]{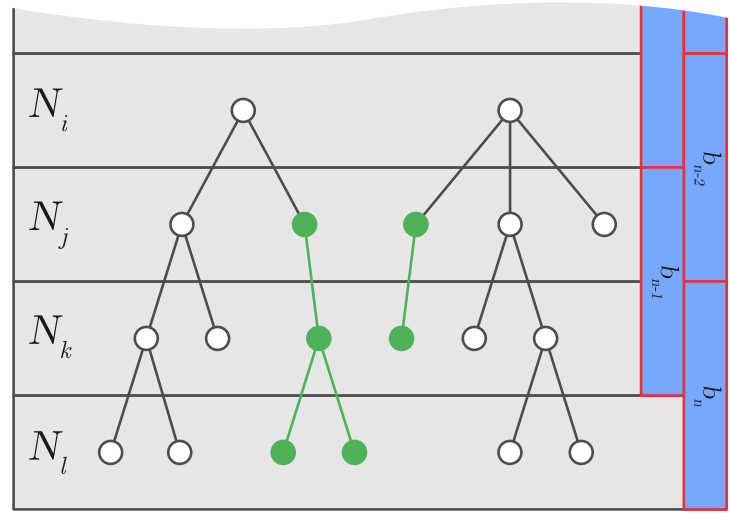}
\caption{Data Provenance of Contaminated Units: As in Figure \ref{fig:EcUND}. blocks are depicted in blue, the connection between blocks (hash pointers) in red. Node $N_i$ is a supplier of raw materials. In two instances these have been contaminated by the mismanagement of $N_j$. Contaminated units are green. The principle of provenance enables latter nodes, including $N_l$ \& $N_k$, to trace the origin of the contaminated goods back up the tree to identify their source.}
\end{figure}

\subsubsection{Provenance of Data}

Theoretically to spend a Bitcoin requires that its provenance be explicitly verified against the entire transaction history of the network from the present epoch through to inception. 
This feature is beneficial to supply chain systems concerned with targeted recall of defective products, especially so since each individual Bitcoin exists as a unique unit within the closed system.
There is no individual actor with the capacity to \texttt{Crtl + c, Ctrl + v} a Bitcoin into existence. 
Bitcoin units are recorded by the Unspent Transaction Output (UTXO) set. 
The balance of any one wallet is the summation over the UTXO instances assigned to the private key with which it is associated.
Analogously an automobile, or similarly a simple chair, is the summation over it's constituent (unique) parts. 

Targeted recall of products effected by particular contagion is an important concern to many organizations, for instance large automotive manufacturers. 
In 2015 the Volkswagen emissions scandal (VW-Abgasaff{\"a}re) prompted a vast recall campaign of large swaths of vehicles, likely including vehicles unaffected by the defective component. 
The problem of tracking unique product constituents through the inherent ``mixing'' process that goes on throughout a supply chain, from material aggregation to finished product, can be conceived of as a task related to that of tracking ``tainted'' coins through the Bitcoin network.
This traceability would allow manufacturers to implement targeted recalls with surgical precision, a substantial efficiency gain for supply chain management systems. 
 
\begin{principle}
Representation of information interchange units should be unique facilitating a navigable trail of provenance for individual components throughout the shared database. 
\end{principle}

\subsubsection{Proof-of-Work}

The Bitcoin proof-of-work exhausts computational power, and ultimately electricity (among other considerations, i.e. the raw materials used to fashion hardware) in the expending of scarce resources, viz. time and money, in order to bring new Bitcoins into existence. 
Reputation and social capital are likewise a scarce resource.
Proof-of-work describes the procedure whereby nodes exchange one resource to be remunerated in kind with another. 
For instance in a supply chain management system nodes consent to the veracity of a transaction by affixing it with their digital signature, expending reputation, and are remunerated with a certified representation of data they consider important. 

\begin{principle}
\label{designp_v}
Transaction quora can establish the degree to which they are concerned with the integrity of some exchange unit by committing to the data structure that represents it with a proof-of-work.
\end{principle}

\section{Threat Model}
\label{sec:greetings}

The Bitcoin protocol is optimized to combat problems inherent to the distributed exchange of value in a peer-to-peer network such as double spending. 
In a supply chain the idea of double spending is nonsensical.
Nodes that maintain a working business relationship preserve this arrangement by acting (in most cases) with integrity. 
Thereby the proposed system is construed to utilize this degree of mutual cooperation between transacting parties, for instance in the exchange quora mechanism (Design Principle \ref{designp_v}). 
This assumption obviates the need for an external arbitrator under the belief that parties at loggerheads would be unable to achieve mutual agreement. 

The sale of counterfeit products is an issue that modern brands with multinational supply chain networks perpetually combat. 
Disingenuous goods circulate widely on the deep web markets mentioned above.
Initiatives such as \texttt{code.moncler.com} by French luxury goods manufacturer Moncler attempts to encourage users to register the QR code stitched into their garments online. 
The system we propose here would enable such manufacturers to keep stricter account of the flow of goods and production materials.
This process is closely related to that of tracking the provenance of contaminated or malfunctioning goods as depicted in Figure 2. 
The simplest potential threat is that of incidental data-corruption, loss, or human-input error. 
The distributed nature of the data model, together with a pre-arranged consensus threshold would serve to stem the adverse effect of this inevitability.

\section{Discussion \& Conclusion}

It has been stated that a ``private blockchain'' is nothing more than an atypical name for a shared database.
In this work we have demonstrated that this is in fact the case. 
However we have also endeavored to provide examples whereby a shared database with certain properties, under specific assumptions, can solve useful problems.
In assessing the merits of a technology one is never fully correct (or fully incorrect) to prefer one method over another.  
In creating this paper we might have individually type-set each letter, printed it with ink, and scanned the result into a computer. 
We chose the more difficult method and used \LaTeX.
In the modern world businesses are often are burdened with long and convoluted supply chains.
The final determination of the degree to which the data management techniques described above are practically useful is empirical. 
In \cite{narayanan2016bitcoin} it is pointed out that Satoshi was probably not an academic because he implemented his system first and wrote about it later.
This work asserts the plausibility of that conjecture insofar as the protocol herein described remains to be actualized. 
What we have done in this work is to venture a rough framework and design methodology. Through careful consideration of the processes by which one might derive concrete value through the actionable properties inherent in the Bitcoin protocol we seek to solve useful problems for the supply chain management community. 














 


%
%

\end{document}